\documentstyle[aps]{revtex}
\input{epsf.tex}
\begin{document}
\draft
\title{Threshold of molecular bound state and BCS transition in dense 
ultracold Fermi gases with Feshbach resonance}
\author{R. Combescot}
\address{Laboratoire de Physique Statistique,
 Ecole Normale Sup\'erieure*,
24 rue Lhomond, 75231 Paris Cedex 05, France}
\date{Received \today}
\maketitle

\begin{abstract}
We consider the normal state of a dense ultracold atomic Fermi gas in 
the presence of a Feshbach resonance. We study the BCS and the 
molecular instabilities and their interplay, within the framework of a 
recent many-body approach. We find surprisingly that, in the 
temperature domain where the BCS phase is present, there is a non zero 
lower bound for the binding energy of molecules at rest. This could 
give an experimental mean to show the existence of the BCS phase 
without observing it directly.
\end{abstract}

Experimental progress in the study of ultracold Fermi gases has 
proceeded quite recently at a very fast pace. Working in the vicinity of a 
Feshbach resonance which allows to cover a very wide range of 
scattering length by sweeping the magnetic field across the resonance, 
several groups dealing either with $ ^{40}$K  \cite{jin,jin1} or $^6$Li 
\cite{slm,grim,hul} have been able to vary the interatomic interaction in 
such a way that the system goes essentially from a weakly attractive 
atomic Fermi gas to a dilute gas of diatomic molecules. The clear 
observation of long lived molecules is one of the very positive 
outcomes of these experiments. More generally these experiments have 
shown that it is quite easy to shift rapidly the magnetic field and thereby 
modify the scattering length $a$, and consequently interactions, in such 
a way that the system does not have time at all to adjust to this change. 
Hence it is experimentally feasible to prepare the gas in an out-of-
equilibrium situation, then to study its evolution from a metastable state 
and in particular the manifestation of various instabilities.

In this paper we study, throughout the $a-T$ phase diagram, the two 
instabilities which arise in a normal Fermi gas due to an attractive 
interaction, namely the molecular and the BCS instabilities. We have 
just stressed that this situation is quite relevant experimentally. We find 
an unexpected interplay between these two instabilities. A particular 
consequence is that, at temperatures where the BCS phase is present, 
there is a nonzero threshold for the binding energy of molecules at rest. 
In other words it should not be possible to observe such a molecule 
with zero binding energy, in contrast to the standard situation for two 
atoms in vacuum where the binding energy of the molecule is zero right 
at the resonance $ a ^{-1}= 0$. Hence quite unexpectedly this link 
between BCS and molecular properties would provide a signature of the 
presence of the BCS phase just by looking at the molecular 
spectroscopic properties, which could be easier to observe 
experimentally than the BCS phase itself.

Actually we have quite recently considered the effect of the Fermi sea 
on the molecular bound state associated with a Feshbach resonance, 
when this is the only instability. We note that, since the Fermi sea is 
responsible for the formation of Cooper pairs, which are some kind of 
molecules, it is somewhat natural that it does also affect the molecular 
properties. We have shown that the presence of the Fermi sea shifts the 
location of the appearance of the molecular state toward positive values 
for the scattering length  \cite{rcnjp}. This can physically be easily 
understood because, if we think of the wavefunction of the (large) 
molecule as made up from plane waves (this corresponds merely to the 
Fourier expansion of the wavefunction), the presence of the Fermi sea 
prohibits the occupation of a number of these plane wave states because 
of Pauli exclusion. So the qualitative effect of this Pauli exclusion is to 
make the building of the molecular state more difficult than in vacuum. 
Naturally this exclusion effect decreases as $T $ increases since the 
statistical occupation of states gets lower. As a result the molecule does 
not appear right at the Feshbach resonance as it occurs in vacuum. 
Instead the inverse scattering length $ a ^{-1}$ has to be larger than a 
positive threshold in order to have an existing molecular bound state. 
To be complete we have to stress that the above picture is for molecules 
with zero momentum for the center of mass. When the molecule has a 
nonzero total momentum, the adverse effect of the Fermi sea will be 
smaller since Pauli exclusion will act on less of the plane wave states 
forming the molecular wavefunction. And it is clear that for a fast 
molecule (with respect to atomic gas velocities) there will be essentially 
no effect of the Fermi sea on the formation of the molecular bound 
state. We note that this shift of the threshold for appearance of the 
molecular states is in qualitative agreement with experimental 
observations of strong losses in $^{6}$Li appearing much below the 
location of the Feshbach resonance \cite{thomas}. However the 
interpretation of these experiments is obviously complex, with in 
particular dynamical effects, and it remains to be seen how strong is 
experimentally the role of the effect that we have just described. The 
best tool is likely to involve some form of spectroscopy.

Here we will use, as in the above study, the result obtained recently  
\cite{rcfesh} for the scattering amplitude due to a Feshbach resonance, 
modified by the presence of the dense Fermi gas. Although obtained 
within a quite general formalism, this result makes use of strong 
simplifications. Nevertheless it is clearly worth exploring simple 
physical approximations before going to more elaborate schemes. First 
the irreducible vertex is taken merely as the simple scattering amplitude 
of two isolated atoms. Second only  Pauli exclusion is taken into 
account for the effect of the Fermi gas. Nevertheless the resulting 
scattering amplitude is quite non trivial. In particular, in contrast with 
the vacuum case, it depends on the total momentum of the scattering 
atoms because of Pauli exclusion. As already mentionned this effect is 
quite small for two atoms with very high momenta, and we will 
concentrate here on the case where the total momentum is zero since it 
displays the strongest manifestation of Pauli exclusion. Also any 
background scattering is omitted for simplicity

The expression \cite{rcfesh} for the full vertex $ \Gamma ( \omega) $ 
in the particle-particle channel, which, except for a factor $ \pi /2 k_{0} 
$, is just the inverse $f ^{-1}$ of the effective scattering amplitude, is 
given by (we take $\hbar=1$):
\begin{eqnarray}
- \frac{2 \pi  ^{2}}{m k _{0}} \frac{1}{ \Gamma ( \omega)} = 
\frac{1}{ \lambda } - \frac{\bar{ \omega }}{ \bar{ W}} + I(\bar{ 
\omega })  \hspace{2cm}
I(\bar{ \omega }) = \int_{0}^{ \infty}dx \, [  1 - \frac{x ^{2}}{ x ^{2} 
- \bar{ \omega }/2} \, \tanh \frac{ x ^{2}-1}{2 \bar{t}}]  
\label{eq1}
\end{eqnarray}
We have taken the chemical potential $ \mu $ (assumed to be positive) 
as our energy scale and our wavevector scale $ k _{0}$ is defined by $ 
\mu = k ^{2}_{0}/ 2m $ (it reduces to the Fermi wavevector at $ T = 0 
$). We have introduced the reduced wavevector $ x = k / k_{0}$, the 
reduced energy $ \bar{ \omega }= \omega / \mu $ and the reduced 
temperature $  \bar{ t }= T  / \mu $. In contrast with Ref. \cite{rcfesh}, 
the origin for the energy $ \omega $ of the two atoms has not been 
shifted at the chemical potential, but merely taken as usual at the bottom 
of the continuous energy spectrum for free particles. The coupling 
constant $ \lambda $ is related to the scattering length $a$ by $ \lambda  
= - 2 k_{0} a / \pi  $. We have introduced the reduced width of the 
Feshbach resonance $  \bar{ W} = m ^{2}  |w| ^{2} / \pi ^{2} k_{0} $ 
where $w$ is the matrix element  \cite{rcfesh} corresponding to the 
Feshbach coupling between the open and the closed channel. In the 
strongly explored case of $ ^{6}$Li , $  \bar{ W}$ is quite large and in 
Eq.(1) we will neglect $\bar{ \omega }/ \bar{ W}$ in the following.

As it is well known the appearance of the molecular bound state will 
show up as a pole in the vertex, and therefore it will be found by 
writing that the r.h.s. of Eq.(1) is zero. We note that the BCS instability 
itself appears also as such a pole. The imaginary part of Eq.(1) is zero 
at the chemical potential $ \omega  = 2 \mu $ and writing that the real 
part is zero leads to:
\begin{eqnarray}
\frac{1}{ a} = \frac{2}{ \pi }  \int_{0}^{ \infty}dk  \, [  1 - \frac{k 
^{2}}{ k ^{2}-2m \mu  } \,  \tanh \frac{ k ^{2}-2m \mu }{4mT}]
\label{eq2}
\end{eqnarray}
which is just the standard BCS equation with our notations (notice that 
$-1/ \lambda $ is changed into $-1/ \lambda + 2/ \bar{ W}$ if we take 
the $  \bar{ W}$ term into account).

It is easily seen from Eq.(2) that the BCS transition is also found in the 
region $ a >0$, that is beyond the location of the Feshbach resonance in 
vacuum. This situation has already been considered by Milstein et al. 
\cite{milstein} and by Ohashi and Griffin \cite{ohgr} within a 
phenomenological fermion-boson model, and it is actually a clear 
ingredient of the BCS-BEC crossover. It seems at first surprising to see 
BCS pairing for $a > 0$ since in this case one has an effective repulsion 
between atoms. But one has rather to consider that the tendency to form 
a bound state is an increasing function of $ a ^{-1}$ (taken  
algebraically), so that going to $ a >0$ makes it even easier to form 
BCS pairs. On the other hand BCS pairing occurs in the Fermi sea : the 
BCS pole appears for a positive energy $ \omega = 2 \mu >0 $. Hence 
the BCS transition stops when $ \mu = 0$. Leggett \cite{legg} has 
already pointed out that, for the excitation spectrum at $T = 0$, there is 
a qualitative change when one crosses $ \mu = 0$. In our case we are in 
the normal state and, at the level of our approximation, the chemical 
potential is merely related to the one-species atom number $n$ by the 
free particle relation:
\begin{eqnarray}
n = \frac{1}{2 \pi ^2} \int_{0}^{ \infty}dk  \, \frac{ k ^{2}}{e ^{(k 
^{2}/2m - \mu )/T} + 1}
\label{eq3}
\end{eqnarray}
Instead of $n$, we introduce for convenience the Fermi wavevector 
defined by $ n = k_F ^{3}/6 \pi ^2$ and the corresponding energy $ 
E_F =  k_F ^{2}/2m$. From Eq.(2) and Eq.(3) we find the critical 
temperature $T_c$ as a function of the scattering length $a$. In Fig. 1 
we plot $ T_{c}/ E_F$ as a function of $ 1/ k_F a$. The end point 
$(T_0 , a_0 ) $ is found at $ T_0/ E_F \simeq 0.99$ and $ 1/ k_F a_0 
\simeq 0.68$.

When $ a ^{-1}$ is increased beyond this point we expect to find an 
instability corresponding physically to a molecular bound state. 
Naturally this is what happens but there is a qualitative change in this 
case. Indeed these molecular states correspond to negative values of the 
energy $ \omega $, for which the r.h.s. of Eq.(1) is always real so we 
are no longer forced to require that its imaginary part is zero. Hence we 
do not need to take a specific energy, and for a fixed temperature $T$, 
we expect to find the binding energy $| \omega |$ increasing with $ a 
^{-1}$, just as it is in the absence of the Fermi sea. Actually this is the 
case that we have explored recently \cite{rcnjp} for $ \mu < 0$, which 
means $ T > T_0$. As already indicated we have found that, due to the 
presence of the Fermi sea, the molecular state with zero binding energy 
does not form for $ a ^{-1} = 0$ as in vacuum, but rather for a positive 
value of $ a_m ^{-1}(T)$ which increases when $T$ decreases. 
Naturally when for fixed $T$ we increase $ a ^{-1}$ beyond $ a_m ^{-
1}(T)$, the molecular binding energy increases from zero. Conversely 
we can say that, starting from large and positive values of $ a ^{-1}$ 
and decreasing it, molecules start to dissociate at $ a_m ^{-1}(T)$. We 
have plotted $ 1/ k_F a_m (T)$ in Fig. 1 and we see that it meets the 
above curve for the BCS transition at the end point. This is natural since 
this point corresponds to cancel the r.h.s. of Eq.(1) for $ \omega =0$ 
and $ \mu = 0$, so it belongs to the two curves. Hence there is a clear 
link between the molecular shift and the existence of the BCS phase.
\begin{figure}
\centering
\vbox to 65mm{\hspace{-6mm} \epsfysize=65mm 
\epsfbox{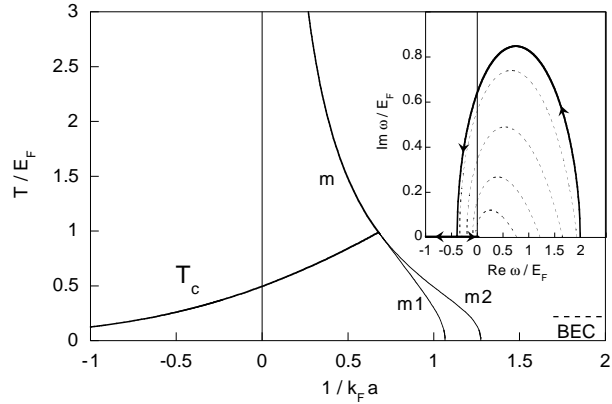} }
\caption{Critical temperature $T_c $ of the BCS transition as a function 
of the inverse scattering length $a^{-1}$, threshold $ a_m ^{-1}(T)$ 
for formation of zero momentum molecules (curve labelled $m$), 
thresholds $ a_{m1} ^{-1}(T)$ (labelled $m1$) and $ a_{m2} ^{-
1}(T)$ (labelled $m2$) (see text). Insert: trajectory of the BCS pole in 
the upper complex $ \omega $ plane for $T=0$ (full curve) and for 
$\bar{t}= 0.2 , 0.5 , 1., 2.$ (successive dashed curves). The arrows 
indicate how the pole(s) moves when $a ^{-1}$ is increasing.}
\label{figure1}
\end{figure}
\vspace{-2mm}
It is now interesting to study in the same way what happens when we 
increase $ a ^{-1}$ at fixed $T$ for $ \mu > 0$, that is $ T < T_0$. 
More specifically we will follow the pole of $ \Gamma ( \omega)$ 
which gives rise to instability. When, for fixed $T$, we start from $ a 
^{-1}$ large and negative there is no instability, until by increasing $ a 
^{-1}$ (algebraically) we reach the BCS instability when we touch the 
$ T_{c}(a ^{-1})$ curve. As it is well known this means that the 
corresponding pole, which was in the lower $ \omega $ complex plane 
has moved up and has reached the real $ \omega $ axis. This is easily 
checked on Eq.(1). More generally, since $ a ^{-1}$ is real, we find 
the trajectory of the pole by writing that the imaginary part in Eq.(1) is 
zero, that is  Im$I(\bar{ \omega })=0$ (if the  $\bar{ W}$ term is 
neglected). The result for this trajectory in the $ \omega $ complex 
plane is shown in the insert of Fig.1 for various temperatures.

We see naturally that, after crossing the real axis, the pole goes deep 
into the upper complex plane, corresponding to the fact that, when we 
enter more into the BCS phase domain, the normal state becomes more 
and more unstable. However when  $ a ^{-1}$ is further increased, the 
pole goes back toward the real $ \omega $ axis, eventually reaching it 
on the semi-axis $ \omega < 0$. This happens for a positive value $ 
a_{m1} ^{-1}(T)$ shown in Fig. 1. At this stage the normal state has 
no longer, strictly speaking, a BCS type instability. It is rather a 
molecular type instability, characterized by a pole on the negative 
energy axis. However what happens next to the pole is rather 
surprising. When $ a ^{-1}$ is increased beyond $ a_{m1} ^{-1}(T)$, 
the pole splits into two poles, one with increasing $ \omega $, the other 
one with decreasing $ \omega $. This is most easily seen at $ T=0$. In 
this case the integration in Eq.(1) is easily performed and one finds for 
the position of the pole on the negative $ \omega $ axis:
\begin{eqnarray}
- \frac{1}{\lambda } \equiv   \frac{\pi }{2k_F} a ^{-1} = 2 r \arctan r  - 
\frac{\pi }{2} r  + 2
\label{eq4}
\end{eqnarray}
where we have set $ r = (|\bar{ \omega }|/2) ^{1/2}$. Starting from $2$ 
at $r=0$, the r.h.s. of Eq.(4) first decreases, reaches a minimum $ -1 / 
\lambda _{min} = 1.67$ for $ \bar{ \omega }= - 0.39 $ ($r=0.30$), 
then increases and behaves asymptotically as $ (\pi / 2) r $. The 
minimum gives the value of $ a_{m1} ^{-1}(T=0)$ for which the BCS 
pole reaches the negative $ \omega $ axis at $ \bar{ \omega }= - 0.39 $. 
When $ a ^{-1}$ is increased beyond $ a_{m1} ^{-1}(0)$, one finds 
two solutions for $ \bar{ \omega }$ around the minimum. One of them 
decreases, in agreement with the expected increase of a molecular 
binding energy. But surprisingly the other one increases and goes 
rapidly to $ \omega =0$, where it disappears (more precisely it goes in 
the lower complex plane along the negative $ \omega $ axis, where it 
has no direct physical manifestation). This disappearance occurs for $ -
1 / \lambda = 2$. We call $ a_{m2} ^{-1}(T=0)= 4 k_F/ \pi $ the 
corresponding value of $ a ^{-1}$. The lower pole is then at $ 
\bar{\omega} = -2.$ Beyond $ a_{m2} ^{-1}( 0)$ there is only one 
solution, corresponding to the continuation of the lower energy 
solution. Hence we recover the standard molecular situation of a single 
pole with a binding energy increasing with $ a ^{-1}$.

In order to understand physically this strange situation it is worthwhile 
to note that, in the absence of a Fermi sea, the r.h.s. of Eq.(4) would 
merely be $ (\pi /2) r $, leading to the usual result $ |\omega | = 1/ma^2$ 
for the molecular binding energy. Hence the Fermi sea contribution is $  
2 r \arctan r  - \pi  r  + 2 = 2  \int_{0}^{1} \! dx \, x ^{2}/( x ^{2}+ r 
^{2})$. Therefore the non analytic decrease of the r.h.s. of Eq.(4), 
behaving as $2 -  \pi (|\bar{ \omega }| /8)^{1/2}$ near $ \bar{ \omega 
}=0$, is produced by the presence of the Fermi sea, and this is this 
decrease which creates the minimum and is responsible for the existence 
of the two poles. This is specifically produced by the $- \pi  r$ term. 
Now this term is easily linked to the density of states for positive 
energy, which is proportional to $ \omega ^{1/2}$. Indeed if we 
evaluate the above integral for small and positive $ \omega  $ (with 
infinitesimal positive imaginary part), we find $  2 R \ln[(1-R)/(1+R)]   
+ i \pi  R  + 2$ with $ R = (\bar{ \omega }/2) ^{1/2}$, and the r.h.s. of 
Eq.(4) is obtained from the analytical continuation of this last result 
through the upper complex plane toward the negative $ \omega $ axis. 
Actually all the information about this Fermi sea contribution is 
contained in the imaginary part $ i \pi  R $ (essentially the density of 
states) since the real part can be recovered through Kramers-Kronig 
relations. Now we note that, for small $ \omega $, this imaginary part 
is positive in contrast with the result Im$I(\bar{ \omega }) = - i (\pi /2) 
R$ for free particles. This gives \cite{rcfesh} for the overall result 
Im$I(\bar{ \omega }) =  i (\pi /2) R$. 

Clearly the physical interpretation of this change of sign is that, below 
the Fermi energy, we deal with hole-like excitations while above the 
Fermi energy we have the standard particle-like excitations: in order to 
create excited states below the Fermi energy we have to remove 
particles since Pauli exclusion forbids to add them. This leads to the 
conclusion that the bound state with very small binding energy we have 
found above, just detached from the Fermi sea, has a hole-like 
character. In other words it corresponds to a molecular state formed by 
two atomic holes (absence of atoms) rather than to a standard molecule 
formed by two atoms. It is a kind of antimolecule. We can come to this 
conclusion more rapidly by thinking of the situation we would have for 
a Fermi energy going to infinity. All the low energy states we could 
consider other than the ground state would be hole-like, since the only 
thing which could be done would be to remove particles. In particular 
we would have only hole-like molecules. Naturally we interpret the 
lower energy pole as a standard particle-like molecular state, since in 
particular it will be the only one present for $ a ^{-1} > a_{m2} ^{-1}( 
0)$. On the other hand, if we think of decreasing  $ a ^{-1}$ below $ 
a_{m2} ^{-1}( 0)$, we will have two poles, one being particle-like and 
the other one hole-like. Therefore when they merge at $ a_{m1} ^{-1}( 
0)$ the corresponding pole has necessarily a mixed particle-hole nature. 
But this is rather natural since this pole is a BCS-type pole as we have 
seen. Indeed it is well known that the BCS instability has this mixed 
nature (we can think of it as the formation of pairs of particles, or as 
well as pairs of holes), as it is reflected in the nature of the low energy 
single particle excitations in BCS theory. Hence it is rather natural that 
when the BCS pole, after moving in the upper complex plane, reaches 
the negative $ \omega $ axis, it has still this mixed nature. This makes 
also likely that, when the two poles separate, they have actually also a 
mixed nature, with a full particle or hole character only reached at $ 
a_{m2} ^{-1}$.

The experimental observability of such a hole-like molecule is unclear. 
Indeed a hole around zero energy is quite deep below the Fermi energy 
and corresponds to a high energy excited state. It is likely that the 
lifetime of such a state will be quite short due to particle-hole 
recombination processes. These would appear because of the strong 
interactions of the atoms in Fermi sea, but they are not present in our 
approximate treatment since we evaluate $I(\bar{ \omega })$ without 
taking into account these interactions. We would have to go beyond this 
theoretical level to make these processes appear. Similarly it is likely 
that the molecular states with partial hole-like nature will have a too 
short lifetime to be directly observed. Hence the physical situation for $ 
a_{m1} ^{-1} < a ^{-1} < a_{m2} ^{-1}$ is quite unclear and it is not 
obvious how this domain will survive in improved theories. In 
particular the existence of two poles may suggest a more complex 
instability, or a forbidden domain with phase separation. On the other 
hand for $ a ^{-1} > a_{m2} ^{-1}$ we are back to a simple physical 
situation. We have just a single pole corresponding to the formation of 
standard molecular state. However the remarkable point is that the 
binding energy $ |\bar{ \omega }|$ is always larger than 2 (i.e. $ | 
\omega |  > 2 E_F$). This means that it is impossible to observe a 
molecular state with zero binding energy, in contrast to the classical 
case of a dilute gas for which the binding energy is zero at the Feshbach 
resonance $ a ^{-1}=0 $. Even if we believe it is possible to observe 
short lived molecules in the range $ a_{m1} ^{-1} < a ^{-1} < a_{m2} 
^{-1}$, this result remains valid as we have seen above. The lower 
bound for the molecular binding energy will just be smaller. Finally we 
have for simplicity limited our explicit quantitative study to the $ T = 0 
$ case, but it is clear from the insert of Fig. 1 that the same results will 
be qualitatively valid for all the range $ 0 \le T \le T_0 $. The lower 
bound for the molecular binding energy will decrease with increasing 
$T$ and go to zero for $ T = T_0 $. Since $ T_0$ is also the maximum 
temperature for the existence of the BCS phase, we reach the surprising 
conclusion that, whenever the BCS phase is present, we can not 
observe a molecule with zero binding energy. This offers an indirect 
way to demonstrate the presence of the BCS condensation.

We have naturally to be quite specific with respect to the above 
statement. In this paper we have only studied the instabilities arising in 
a normal Fermi gas, with in particular no molecules already present. 
Therefore when we consider this gas for $ a ^{-1} > a_{m2} ^{-
1}(T)$, we deal with an out of equilibrium situation since at equilibrium 
a sizeable fraction of the gas should be under molecular form. Therefore 
we have in mind an experiment where, starting from the $ a ^{-1} < 0$ 
side of the Feshbach resonance, one would very rapidly change $ a ^{-
1} $ by acting on the magnetic field and then observe the binding 
energy of the first few molecules appearing in the gas. We could also 
worry about the effect of the Bose-Einstein condensation of molecules. 
However in our simple picture the critical temperature for this BEC is 
\cite{milstein} $ T _{BEC}= 0.218 T_F$, so the effect we have 
considered could be at least observed in the range $ T _{BEC} < T < 
T_0 $.  Also the instabilities we have considered are all at zero 
wavevector, just as the dominant BCS instability itself. So the 
molecules we considered have zero total momentum. For nonzero 
momentum the effects will be weaker, and we expect them to disappear 
at some wavector, whose inverse is of order of the Cooper pair size, 
just as the BCS instability itself.

Finally we have naturally to consider that our theoretical approach is 
clearly not quantitatively accurate, since for example our value for the 
BCS critical temperature $T _{c}$ is just the standard one 
\cite{stoofal}, and does not contain lower order fluctuation effects 
\cite{gork} nor higher orders and self-energy effects \cite{rc}. 
However we believe that our results remain qualitatively valid under 
much more general conditions. Indeed we see from Fig. 1 that the 
nonzero threshold for molecular binding energy that we have found is 
ultimately linked to the trajectory of the BCS pole in the upper complex 
plane. But the existence of a BCS pole, and its trajectory in the upper 
complex plane are quite general features of any theoretical description. 
Since we naturally expect to find molecules for very large $ a ^{-1}$, 
this pole has to go back to the real negative $ \omega $ axis, as we have 
found, which implies again the nonzero molecular binding energy 
whenever the BCS phase is present. In other words the (qualitative) 
topology of our results should remain valid even if they are 
quantitatively modified.

We stress again that what we have done in this paper is to look at the 
instability properties of the normal state, even in regions where it is not 
the equilibrium state. We have found that, when $ a ^{-1}$ goes from 
$ - \infty $ to $ + \infty$ across the phase diagram, these properties do 
not change smoothly, but rather that the behaviour changes on some 
lines that we have identified. One can wonder if the same is not true for 
the equilibrium phase, and in particular for the ground state in the 
superfluid domain for which a quite natural interpolating hypothesis 
\cite{legg,nsr} is a BCS-like ground state whatever the scattering 
length, since this wavefunction is known to describe properly the dilute 
molecular ground state \cite{kk} as well as the weak coupling BCS 
state. In this case one would rather expect a single instability to appear 
corresponding to the pair formation, so it might be that this BEC-BCS 
crossover is not as smooth as suggested by this picture, and that the 
actual physics is more complex. However our results have no direct 
bearing on this question, since we have explored only the normal 
phase, not the superfluid. We just note that, in the superfluid state, the 
effect of Pauli exclusion will still be present.

We are very grateful to T. Bourdel, Y. Castin, C. Cohen-Tannoudji, J. 
Dalibard, X. Leyronas, C. Mora and C. Salomon for very stimulating 
discussions.

* Laboratoire associ\'e au Centre National
de la Recherche Scientifique et aux Universit\'es Paris 6 et Paris 7.

\end{document}